\documentclass[aps,prl,twocolumn,showpacs,nofootinbib,
superscriptaddress,nobalancelastpage,nobibnotes]{revtex4}
\usepackage{graphicx}
\usepackage{amsmath}
\usepackage{epsfig}
\usepackage{psfig}

\def\beq{\begin{equation}}
\def\eeq{\end{equation}}
\def\beqa{\begin{eqnarray}}
\def\eeqa{\end{eqnarray}}
{\newcommand{\lsim}{\mbox{\raisebox{-.6ex}{~$\stackrel{<}{\sim}$~}}}
{
\def\dstyle{\displaystyle}
 
\def\onethird{\frac{1}{3}} 
 
\def\twothird{\frac{2}{3}} 
\def\onesixth{\frac{1}{6}}
\def\xsixth{\frac{\dstyle x}{6}}
\def\onebyroottwo{\frac{1}{\sqrt{2}}}
\def\cm{\,{\rm cm}}

\def\m{\,{\rm m}}

\def\mpc{\,{\rm Mpc}}

\def\ev{\,{\rm eV}}

\def\gev{\,{\rm GeV}}
\def\tev{\,{\rm TeV}}
\def\pev{\,{\rm PeV}}

\def\nubar{\bar{\nu}}
\def\nue{\nu_e}
\def\numu{\nu_\mu}
\def\nutau{\nu_\tau}
\def\nuebar{\nubar_{e}}
\def\numubar{\nubar_{\mu}}
\def\nutaubar{\nubar_{\tau}}
\begin{document}
\title{\bf Probing neutrino mixing angles with ultrahigh energy neutrino 
telescopes} 
\author{Pijushpani Bhattacharjee}
\email{pijush@iiap.res.in}
\affiliation{Indian Institute of Astrophysics, 
Bangalore-560 034, India}                            
\author{Nayantara Gupta}
\email{nayan@imsc.res.in}
\affiliation{Institute of Mathematical Sciences, C.I.T. Campus, 
Taramani, Chennai- 600 113. India}                            
\begin{abstract}
\noindent 
We point out that detecting $\nuebar$'s from distant 
astrophysical sources with the up-coming and future neutrino telescopes   
using the Glashow resonance  
channel $\nuebar e^{-}\to W^{-} \to$ anything, which
occurs over a small energy window around the $\nuebar$ energy of 
$\sim6.3\pev$, offers a new way of measuring or setting limits on  
neutrino mixing angles, in particular the angle $\theta_{12}$, thereby 
providing an independent experimental probe of neutrino mixing angles.  
We also discuss how this exercise may throw light on the nature of the 
neutrino production mechanism in individual astrophysical sources. 
\end{abstract}



\maketitle
%
Experimental determinations of the parameters governing neutrino flavor 
oscillation phenomena, such as the mass-square 
differences and mixing angles amongst the three 
light neutrino species, have so far been done with solar- atmospheric-, 
reactor-, and accelerator neutrino experiments~\cite{pdg} that involve 
neutrinos of few MeV -- GeV energies. At a purely phenomenological level, 
one may ask if neutrinos of vastly different energies 
oscillate in the same way. In this context it is important to think 
of other possible independent ways of measuring the neutrino parameters 
that involve neutrinos of different (in particular higher) energies than 
those currently employed in neutrino oscillation experiments. 

In this Letter we point out that up-coming and future neutrino telescopes   
capable of detecting neutrinos of very high (at least up to 
several PeV:\ \ $1\pev=10^{6}\gev$) energies, would potentially also be 
able to measure one or more of the neutrino mixing angles at such high 
energies~\cite{fn1}. Specifically, we suggest that detecting $\nuebar$'s 
from distant astrophysical sources through the Glashow resonance 
(GR)~\cite{GR} channel, $\nuebar e^{-}\to W^{-} \to\, $ anything, which
occurs over a small energy window around the $\nuebar$ energy 
$E_{\nuebar}^{\rm GR}=m_W^2/2m_e=6.3\pev$, offers a new way of 
measuring the angle $\theta_{12}$ (see below)~\cite{fn_athar}. 

Neutrino flavor oscillation phenomena are governed by six 
independent parameters\cite{nu_osc_books}: two mass-squared differences, 
$\Delta m^2_{12}\equiv |m^2_2-m^2_1|$ and $\Delta m^2_{23}\equiv 
|\m^2_3-m^2_2|$, three mixing angles, $\theta_{12}\,, \theta_{23}\,,$ 
and $\theta_{13}$, and a possible CP-violating phase $\delta$. 
Here ($m_1\,, m_2\,, m_3$) are the masses corresponding to the three light 
neutrino mass eigenstates ($\nu_1\,, \nu_2\,, \nu_3$). The mixing angles 
parametrize the relation between the mass eigenstates and the flavor 
eigenstates ($\nue\,,\numu\,,\nutau$); see equation (\ref{U_general}) 
below. The current experimental  
situation, as summarized in Ref.~\cite{pdg}, is the following:    
the solar neutrino data are consistent with flavor 
oscillations mainly between $\nue$ and $\numu$ driven by a mass-squared 
difference $\Delta m^2_{\rm solar}=\Delta m^2_{12}\approx 
7.1\times10^{-5}\ev^2$ and 
mixing angle $\theta_{\rm solar}=\theta_{12}\approx 32^{\circ}.5$, while 
the atmospheric 
neutrino data are explained by oscillations mainly between $\numu$ and 
$\nutau$ with $\Delta m^2_{\rm atm}=\Delta m^2_{23}\approx 
2.6\times10^{-3}\ev^2$ and
$\theta_{\rm atm}=\theta_{23}\approx 45^{\circ}$, the maximal value.  
The other mixing angle $\theta_{13}$ is 
constrained by reactor neutrino experiments to be very small, 
$\theta_{13}\lsim 9^\circ$. The CP-violating phase $\delta$ remains 
undetermined by present experiments. The uncertainties (currently roughly 
10--20\%) in the values of the above parameters may be expected to be 
significantly reduced by future experiments. 

Very high energy neutrinos of energies well above several hundreds of TeV, 
and in some cases extending well into EeV ($10^9\gev$) energies, are 
predicted to be produced by astrophysical sources such as 
Active Galactic Nuclei (AGN) and Gamma Ray Bursts 
(GRB)'s~\cite{agn_grb_nu_rev}. More 
``exotic'' sources such as cosmic topological defects are 
predicted to produce neutrinos of energy even up to 
several hundred EeV~\cite{td_nu}. Detection of such high energy 
neutrinos from distant astrophysical sources is the primary goal of a host 
of proposed or already-under-construction large under-ice and underwater 
optical Cerenkov detectors such as ICECUBE~\cite{icecube}, 
ANTARES~\cite{antares}, NESTOR~\cite{nestor}, NEMO~\cite{nemo}, as well as 
radio Cerenkov detectors such as RICE~\cite{rice} and ANITA~\cite{anita}. 
These detectors will be able to determine the energy as well the arrival 
directions of the neutrinos enabling point source neutrino astronomy. 

High energy neutrinos are produced in astrophysical sources mainly from 
the 
decays of charged pions (and kaons) which are produced through interaction 
of high energy protons with either ambient photons (``$p\gamma$'') or 
protons (``$pp$'') within the 
source~\cite{berezinsky_book_plus_gaisser-halzen-stanev-95}. 
In the $p\gamma$ process, charged pion (and hence neutrino) production 
occurs dominantly through production of the $\Delta$ 
resonance~\cite{stecker68}: $p\gamma\to\Delta^+\to\pi^+n$, with 
$\pi^+\to\mu^+\numu$, and $\mu^+\to e^+\nue\numubar$. Thus each $p\gamma$ 
interaction eventually produces one each of $\nue$, $\numu$, and 
$\numubar$. Denoting by ${\mathcal F}_{\nu}\equiv 
\{\nue\,,\nuebar\,,\numu\,,\numubar\,,\nutau\,,\nutaubar\}$ the 
fractional amounts of $\nu$'s and $\nubar$'s of various flavors, we thus 
have for the $p\gamma$ process the flavor fractions at the source,   
${\mathcal F}_{\nu}^{p\gamma}= 
\{\onethird\,,0\,,\onethird\,,\onethird\,,0\,,0\}$. 
In the $pp$ process, on the other hand, each inelastic $pp$ 
collision produces a nearly equal mix of $\pi^+$'s and $\pi^-$'s whose 
decays together produce the flavor fractions, 
${\mathcal F}_{\nu}^{pp}= 
\{\onesixth \,,\onesixth\,,\onethird\,,\onethird\,,0\,,0\}$. Note that 
the $(\nu + \nubar)$ fractions,  
${\mathcal F}_{(\nu+\nubar)}\equiv \{(\nue+\nuebar)\,, 
(\numu+\numubar)\,, (\nutau+\nutaubar)\}=\{\onethird\,,\twothird\,,0\}$ 
for {\it both} $p\gamma$ and $pp$ processes. 

In the tenuous (radiation dominated) environments within 
most potential astrophysical sources such as AGNs and GRBs, the 
dominant production mechanism for high energy neutrinos is expected to be 
the $p\gamma$ 
process~\cite{berezinsky_book_plus_gaisser-halzen-stanev-95}. 
However, for generality let us assume that a  
fraction $x$ of the total number of all $\nu$'s and $\nubar$'s produced at 
the source are of $pp$ origin and the rest of $p\gamma$ origin. Thus, 
$x=0$ for a pure $p\gamma$ origin of the neutrinos. The 
flavor fractions at the source now work out to be   
\begin{equation}
{\mathcal F}_{\nu}^{\rm source}=\left\{
\left(\onethird-\xsixth\right)\,, \xsixth\,, 
\onethird\,, \onethird\,, 0\,, 0
\right\}\,.
\label{source_fracs}
\end{equation}
Note that $(\nu+\nubar)$ fractions of 
different flavors at source are again ${\mathcal F}_{(\nu+\nubar)}^{\rm 
source} = \{\onethird\,,\twothird\,,0\}$, {\it independent} of the 
fraction $x$. 

The flavor fractions of the $\nu$'s and $\nubar$'s arriving at Earth 
are different from the above fractions at the source due to neutrino  
flavor oscillations during the propagation from the source to the Earth. 
In the standard, 3-generation scenario of neutrino oscillations, 
the flavor eigenstates $|\nu_{\alpha}\rangle\; (\alpha=e\,,\mu\,,\tau)$ 
are related to the mass eigenstates
$|\nu_i\rangle\; (i=1,2,3)$ through $|\nu_{\alpha}\rangle = \sum_i 
U_{\alpha i}^*|\nu_i\rangle$. 
For values of $\Delta m^2$ as indicated by solar and atmospheric neutrino 
experiments, the vacuum oscillation length $L_{\rm osc}=4\pi E_\nu/|\Delta 
m^2|\simeq 
2.5\times10^{-12}\left(E_\nu/1\tev\right)\left(10^{-5}\ev^2/|\Delta 
m^2|\right)\mpc$ is always much smaller than the distances (several 
hundreds to thousands of Mpc) to possible astrophysical sources of 
neutrinos such as AGNs and GRBs, even for the highest energies of 
interest. The neutrinos thus oscillate many times before 
reaching Earth, and the oscillation (or survival) probabilities averaged 
over many oscillations take the simple form~\cite{nu_osc_books}
\begin{equation} 
P(\nu_\alpha \to \nu_\beta)=\sum_{i=1,2,3} \mid U_{\alpha i}\mid^2 \;\; 
\mid U_{\beta i} \mid^2\; . 
\label{osc_prob}
\end{equation} 
A ``standard'' form~\cite{pdg} of the neutrino mixing matrix $U$ is 
\begin{widetext}
\begin{equation}
U=\left(\begin{array}{ccc}
U_{e1} & U_{e2} & U_{e3} \cr
U_{\mu1} & U_{\mu 2} & U_{\mu 3} \cr
U_{\tau1} & U_{\tau2} & U_{\tau3}
\end{array}\right)
= \left(\begin{array}{ccc}
c_{12}c_{13} & s_{12}c_{13} & s_{13}e^{-i\delta} \cr
-s_{12}c_{23} - c_{12}s_{23}s_{13}e^{i\delta} & c_{12}c_{23} -
s_{12}s_{23}s_{13}e^{i\delta} & s_{23}c_{13} \cr
s_{12}s_{23} - c_{12}c_{23}s_{13}e^{i\delta} &  -c_{12}s_{23} -
s_{12}c_{23}s_{13}e^{i\delta} & c_{23}c_{13}
\end{array}\right)\times {\rm 
diag}\left(e^{-i\varphi_1/2}\,,e^{-i\varphi_2/2}\,,1\right)\;,
\label{U_general}
\end{equation}
\end{widetext}
where $s_{ij} = \sin\theta_{ij}$, $c_{ij} = \cos\theta_{ij}$, and 
$\delta\,, \varphi_1\,,$ and $\varphi_2$ are CP-violating phases. The 
phases $\varphi_1$ and $\varphi_2$ are present only for Majorana 
neutrinos; they do not enter into the expressions for 
oscillation probabilities~\cite{nu_osc_books}, although the ``Dirac'' phase 
$\delta$, if it is non-zero, does. 

In our numerical calculations below we use the 
above full form of the mixing matrix (\ref{U_general}) in calculating 
the expressions for various oscillation probabilities (\ref{osc_prob}) 
with $\delta$ set to zero for simplicity. However, 
simple analytical derivation of our main results are 
possible if we set $\theta_{13}=0^\circ$ and $\theta_{23}=45^\circ$ 
(conditions which we hereafter refer to as ``optimal''), in 
which case the mixing matrix $U$ takes the simple form 
\begin{equation}
U_{\rm{optimal}} = \left(
\begin{array}{ccc}
c_{12} & s_{12} & 0 \\ 
-\onebyroottwo s_{12} & \onebyroottwo c_{12} & \onebyroottwo \\ 
\onebyroottwo s_{12} & -\onebyroottwo c_{12} & 
\onebyroottwo
\end{array}
\right) \;,
\label{U_optimal}
\end{equation}
involving only the mixing angle $\theta_{12}$. 

Using Equations (\ref{source_fracs}), (\ref{osc_prob}), 
and (\ref{U_optimal}) we can write the 
flavor fractions of $\nu$'s and $\nubar$'s arriving at Earth in the 
optimal case as 
\begin{equation} 
{\mathcal F}_{\nu\;,{\rm optimal}}^{\rm 
Earth}=\left\{\left(\onethird-p\right)\,,p\,,\frac{1+q}{6}\,,\frac{1-q}{6}\,, 
\frac{1+q}{6}\,,\frac{1-q}{6}\right\}\,,
\label{earth_fracs}
\end{equation}
where 
$p=\onesixth\left(x+2\xi-2x\xi\right)\,$, $q=\left(\xi-x\xi\right)\,$, 
with $\xi\equiv s_{12}^2c_{12}^2=\frac{1}{4}\sin^2 2\theta_{12}$. 

Note that $(\nu+\nubar)$ fractions of different flavors at Earth are 
simply $\{\onethird\,,\onethird\,,\onethird\}$, {\it independent} of the 
fraction $x$ in the optimal case. Also, the 
$(\nu+\nubar)$ fractions are independent of the neutrino mixing angles in 
the optimal case. Dependence on $x$ and mixing angles (only $\theta_{12}$ 
in the optimal case) is contained in the 
separate number fractions of $\nu$'s and $\nubar$'s of individual flavors. 
However, separate identification of the $\nu$'s and 
$\nubar$'s of individual flavors is not generally possible through the 
usual charged-current (CC) interactions of neutrinos in the 
currently operating or planned water or ice-based detectors, since the 
sign of the charge of the produced charged lepton cannot be determined.  
The only exception is the $\nuebar$ which can be identified 
in water or ice based detectors through the Glashow 
Resonance (GR) channel already mentioned~\cite{fn2}. An 
experimentally measurable ratio such as $\nuebar/(\numu+\numubar)$ giving 
the $\nuebar$ fraction to $(\numu+\numubar)$ fraction measured over an  
energy interval centered at the GR 
energy can then yield a measurement of the angle $\theta_{12}$. 
Note that taking the ratio as above cancels out 
the unknown total neutrino flux coming from the source. Alternatively, 
if the spectrum of the neutrinos is the same across a broad range of 
energies, then one can also get the absolute normalization of the total 
neutrino flux from the total number of $(\numu+\numubar)$ 
events over a suitably chosen range of energies. 

The cross section for the GR interaction of $\nuebar$ can be written 
as~\cite{GR,anchor_hepph_0410003}
\beq
\sigma^{\rm GR}\simeq 0.675 
\left(\frac{m_W^2}{2m_eE_{\nuebar}}\right)\delta\left(2m_eE_{\nuebar} 
- m_W^2\right)\,,\label{GR_X_section}
\eeq
where $m_e$ and $m_W$ are the $e^{-}$ and $W$ masses. The $\numu$ CC 
interaction cross section 
is~\cite{gandhi_etal} $\sigma_{\numu,\numubar}^{\rm CC}\simeq 
5.53\times10^{-36}\cm^2(E_\nu/\gev)^{0.363}$. Integrating the flux times 
cross section over an energy bin of say  
$5.01\pev$ -- $7.9\pev$ (which spans 0.1 in $\log_{10}$ on 
either side of the resonance energy 6.3 PeV), we can 
calculate the {\it ratio}, $R\equiv \nuebar^{\rm GR}/(\numu+\numubar)^{\rm 
CC}$, of GR to $(\numu+\numubar)$-CC event rates in a water or ice based 
detector. For an assumed $E_\nu^{-2}$ neutrino spectrum we get 
in the optimal case, 
\beq
R_{\rm optimal}=30.5 
\left[\xi+\frac{x}{2}\left(1-2\xi\right)\right]\,.
\label{ratio_eqn}
\eeq 
In obtaining equation (\ref{ratio_eqn}) we have considered muons with 
contained vertices only~\cite{beacom_etal_prd68_093005_2003}.

\begin{figure}
\includegraphics[width=3.25in]{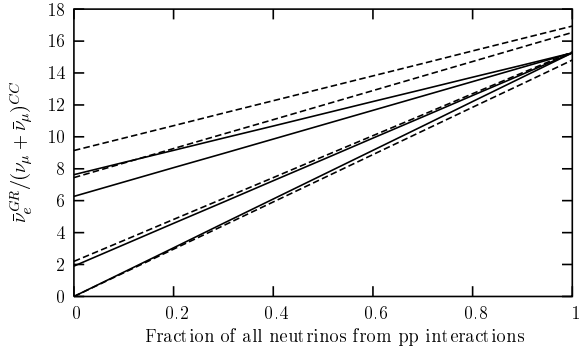}
\caption{\label{fig:GR_numucc_theta_23_var} 
The ratio of number of Glashow resonance events to $(\numu+\numubar)$ 
charged current events in a water or ice based detector as a function of 
the fraction $x$ of $(\nu+\nubar)$'s that originate from 
$pp$ interactions at the source. 
The solid lines, from bottom to top, are for $\theta_{12}=0^\circ\;, 
15^\circ\;, 32.5^\circ\;,$ and $45^\circ$ --- all for the ``optimal'' case 
$\theta_{23}=45^\circ$ and $\theta_{13}=0^\circ$. 
The dashed lines are in the same order of $\theta_{12}$ values as the 
solid lines --- all for $\theta_{13}=0^\circ$, but   
$\theta_{23}=40^\circ$. 
}
\end{figure}

\begin{figure}
\includegraphics[width=3.25in]{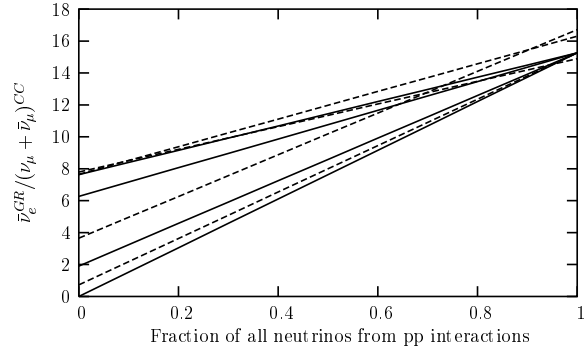}
\caption{\label{fig:GR_numucc_theta_13_var}
Solid lines: same as in Fig.\ \ref{fig:GR_numucc_theta_23_var}. 
The dashed lines are in the same order of $\theta_{12}$ values as the 
solid lines --- all for $\theta_{23}=45^\circ$, but 
$\theta_{13}=9^\circ$.
}
\end{figure}

Equation (\ref{ratio_eqn}), which is a linear function of $x$, is 
displayed in 
Figures \ref{fig:GR_numucc_theta_23_var} and 
\ref{fig:GR_numucc_theta_13_var} 
for various values of $\theta_{12}$ (solid lines). In these Figures we 
also show the (dashed) lines obtained by allowing small deviations of the 
angles $\theta_{23}$ (Fig.~\ref{fig:GR_numucc_theta_23_var}) 
and $\theta_{13}$ (Fig.~\ref{fig:GR_numucc_theta_13_var}) from their 
``optimal'' values; 
this we do by using the general form of the mixing 
matrix given by Equation (\ref{U_general}). 

Figures \ref{fig:GR_numucc_theta_23_var} and 
\ref{fig:GR_numucc_theta_13_var} show that the dependence of the ratio 
$R$ on the mixing angles progressively weakens as $x$ (i.e., fraction 
of $pp$ origin neutrinos) increases and, in the optimal case, disappears 
completely for $x=1$ as expected (see equation \ref{earth_fracs}). 
Fortunately, for the dominant $p\gamma$ process ($x=0$) the 
ratio $R$ has significant dependence on $\theta_{12}$ thus allowing the 
possibility of measuring this angle. 
The flip side is that the expected number of $\nuebar$ GR 
events is smaller in the case of $p\gamma$ as compared to $pp$ dominance.     

Clearly, the accuracy with which the different mixing angles can be probed 
will be limited by the accuracy with which the ratio $R$ can be measured 
in a real detector. While the ratio $R$ always increases with 
increasing $\theta_{12}$, its dependences on small deviations of the 
values of $\theta_{23}$ and $\theta_{13}$ from their optimal values are 
somewhat more complicated, though perhaps too small to be measurable. 
From the above Figures we see that for a given upper limit 
(say $\lsim 10\%$) on the fraction $x$ for a given source from independent 
considerations (e.g., high energy $\gamma$-ray observations), a 
measurement of the ratio $R$ yields a lower limit on the 
angle $\theta_{12}$. Conversely, for  a given fixed value of 
$\theta_{12}$, a measured value of $R$ gives a 
value of the fraction $x$, thus providing important clues to the nature of 
the neutrino production mechanism in the 
source under consideration.  

At PeV energies neutrinos begin to get absorbed in passing through 
the Earth; this is especially true for $\nuebar$ at the GR energy where 
the $\nuebar$ interaction cross section is significantly enhanced. Thus 
only the downward-going to horizontal $\nuebar$'s will be detectable.   
Fortunately, the atmospheric neutrino flux at PeV energies is negligible 
and may not be of concern as a background. The detectability of $\nuebar$ 
GR events has been discussed recently for the diffuse neutrino background 
(due to integrated contribution from all sources in the Universe)  
in Ref.~\cite{anchor_hepph_0410003} for a ICECUBE-type detector. 
The $\nuebar$ GR events can be identified through the detection of the 
electromagnetic shower produced by the hadronic decay products of the $W$ 
in the detector. This corresponds to $\sim$70\% detection 
efficiency~\cite{anchor_hepph_0410003} reflecting the branching ratio of 
hadronic decay of the $W$. The main background, then, are the showers due 
to $(\nue+\nuebar)$-CC events within the GR energy 
window~\cite{anchor_hepph_0410003}. Interestingly, since, at least in the 
optimal case, the $(\nue+\nuebar)$ fraction is equal to the 
$(\numu+\numubar)$ fraction, and since the CC cross section is essentially 
the same~\cite{gandhi_etal} for $\numu$'s and $\nue$'s, our ratio $R$ in 
equation (\ref{ratio_eqn}) multiplied by a factor of $\sim 0.7$ gives a 
good estimate of the expected ratio of signal-to-background number of 
events. At the same time, the accuracy with which the ratio $R$ can be 
measured in a detector depends on the actual number of GR events detected  
in a reasonable time frame (say 6--7 years), which in turn depends on the 
flux from the source and the size of the detector.  

In this context, we should mention that throughout the above discussion we 
have implicitly assumed individual point sources for which the meaning of 
the $pp$ origin fraction $x$ is well-defined. For the diffuse background, 
on the other hand, that fraction refers to some kind of weighted average 
over all sources. Our discussions above are then valid for diffuse 
background as well provided the above difference in the meaning of $x$ is 
kept in mind. Clearly, from the point of view of 
signal-to-background ratio, individual point sources such as not too 
distant and yet sufficiently powerful GRBs, would be easiest to detect 
because then temporal as well as directional information could be used to 
increase the signal-to-background ratio. Also, individual sources are more 
easily subject to observations in other bands such as high energy 
$\gamma$-ray observations which may provide independent information on the 
fraction $x$ (assuming hadronic origin of $\gamma$-rays through $\pi^0$ 
decay). 
In any case, in view of important information regarding 
neutrino parameters as well as the nature of high energy astrophysical 
sources of neutrinos that may be obtained from detection of $\nuebar$ GR 
events as discussed above, it would certainly be useful in designing 
future high energy neutrino telescopes to optimize them for detection 
of possible $\nuebar$'s from astrophysical sources through the GR channel.  

We thank K.~Balaji, R.~Mohapatra, and G.~Rajasekaran for pointing out a 
major error in the motivational basis of an earlier version of this 
paper, which has led to a major shift of emphasis in the present version, 
although the main calculations and results remain the same.  
One of us (NG) wishes to thank IIA, Bangalore for hospitality. This 
work is partially supported by a NSF US-India cooperative research grant 
to PB at IIA, Bangalore. 

%
%


\begin{thebibliography}{99}
\bibitem{fn1} Unlike the mixing angles, the mass-square differences 
($\Delta m^2$) appear only in the oscillatory terms in the flavor 
oscillation probabilities which, for cosmologically distant sources and   
for currently known experimental values of $\Delta m^2$, 
average out to zero. The $\Delta m^2$'s cannot 
therefore be probed with high energy 
neutrinos from distant astrophysical neutrino sources for their 
currently known values.
\bibitem{fn2} Detection of $\nuebar$'s through the GR channel in ICECUBE 
type detectors has been discussed in Ref.~\cite{anchor_hepph_0410003}. 
However, they have not considered the possibility of using this to measure 
the angle $\theta_{12}$.
\end{thebibliography}
\end{document}